\documentclass[pra,superscriptaddress,showpacs,twocolumn]{revtex4}

\usepackage{amsmath,amssymb} 
\usepackage{graphicx,color}
\usepackage{dcolumn}

 \newcommand{\ket}[1]{|{#1}\rangle} 
 \newcommand{\bra}[1]{\langle{#1}|}

\newcommand\ketbra[2]{| #1 \rangle\!\langle #2 |}

\newcommand{\Ham}{\mathcal{H}}

\begin{document}
    
\title{Liquid State NMR as a Test-bed for Developing Quantum Control Methods}

\author{C.A. Ryan\footnote{These authors contributed equally to this work.}}
\affiliation{Institute for Quantum Computing and Dept. of Physics, University of Waterloo, Waterloo, ON, N2L 3G1, Canada.}

\author{C. Negrevergne$^*$}
\affiliation{Institute for Quantum Computing and Dept. of Physics, University of Waterloo, Waterloo, ON, N2L 3G1, Canada.}

\author{M. Laforest }
\affiliation{Institute for Quantum Computing and Dept. of Physics, University of Waterloo, Waterloo, ON, N2L 3G1, Canada.}

\author{E. Knill}
\affiliation{Mathematical and Computational Sciences Division, National Institute of Standards and Technology, Boulder, Colorado 80305, U.S.A.}

\author{R. Laflamme}
\affiliation{Institute for Quantum Computing and Dept. of Physics, University of Waterloo, Waterloo, ON, N2L 3G1, Canada.}
\affiliation{Perimeter Institute for Theoretical Physics, Waterloo, ON, N2J 2W9, Canada}

\begin{abstract}
In building a quantum information processor (QIP), the challenge is to
coherently control a large quantum system well enough to perform an
arbitrary quantum algorithm and to be able to correct errors induced
by decoherence. Nuclear magnetic resonance (NMR) QIPs offer an
excellent test-bed on which to develop and benchmark tools and
techniques to control quantum systems. Two main issues to consider when designing control methods are
accuracy and efficiency, for which two
complementary approaches have been developed so far to control qubit
registers with liquid-state NMR methods. The first applies optimal
control theory to numerically optimize the control fields to implement
unitary operations on low dimensional systems with high fidelity. The
second technique is based on the efficient optimization of a sequence
of imperfect control elements so that implementation of a full quantum
algorithm is possible while minimizing error accumulation. This
article summarizes our work in implementing both of these methods.
Furthermore, we show that taken together, they form a basis to design
quantum-control methods for a block-architecture QIP so that large system
size is not a barrier to implementing optimal control techniques.
\end{abstract}  


\maketitle

\section{Introduction}
Quantum mechanics has been successfully used to describe experimental
phenomena for close to a century. However, it was only much more
recently that the idea of using quantum mechanical evolution to
process information was conceived~\cite{Feynman:1982a}. This is
revolutionizing the way we view the complexity of information
processing. In particular, quantum information processors (QIPs) may
be able to perform certain computational tasks, such as factoring
large integers or simulating quantum systems, exponentially faster
than classical computers \cite{IkeandMike}. One of the difficulties in
applying quantum mechanics to information processing is that quantum
effects are extremely delicate and difficult to detect.   The
system must be well isolated from the environment, or decoherence will
quickly destroy the quantum superpositions and entanglement that the
QIPs exploit and reduce the system to a probabilistic classical state.
It has only recently been realized that it is possible to efficiently encode
quantum information to protect it from decoherence. As a result,
the control and isolation of quantum information systems need not be
arbitrarily good. It suffices for the computationally relevant error
probability per basic quantum gate to be $\lesssim10^{-4}$
\cite{Preskill:1998a} and with extreme overheads $\lesssim10^{-2}$ \cite{Knill:2005a}. In practice, satisfying this error bound requires balancing the needs for strong interactions to be able to
apply quantum gates and good isolation to ensure that the quantum
systems evolve coherently. Moreover, in the context of building a
large-scale QIP, the classical resources required to implement the
desired control must scale efficiently with the size of the QIP. This
is an extraordinary challenge, but the prospect of being able to
exploit the power of quantum computing has provided the impetus to
explore many technologies in the pursuit of a scalable quantum
computer \cite{QIP:2004a}. 

One such technology is based on nuclear magnetic resonance phenomena.
Nuclear magnetic resonance (NMR) phenomena are well described by
quantum mechanics \footnote{Although a contrived classical description
is possible for few spins at low polarizations, see Ref.
\cite{PhysRevLett.88.167901}.}, and the control technology developed
over its more than 50 years of existence has allowed for sophisticated
pulse sequences that exploit the full range of the system dynamics. By
leveraging NMR engineering experience and its well developed control
technology, NMR has been used to realize many of the first
demonstrations of quantum algorithms. 

As NMR quantum information experiments increased in complexity, so did
the quantum control methods used. In the first implementations of
quantum algorithms on small molecules, such as the Deutsch-Jozsa
\cite{Chuang:1998fk} and Grover's \cite{Jones:1998qy} algorithms,
preparation of the three-qubit GHZ state \cite{Laflamme:1998lr} and the
implementation of the three-qubit quantum error correction
code\cite{Cory:1998a}, the control sequences were written entirely by
hand based on well-known techniques from NMR and heuristic optimization.
In later more sophisticated experiments on larger molecules, such as a
seven-qubit cat state experiment and the implementation of the five-qubit code
\cite{Knill:2000a,Knill:2001fk}, the control sequences were still
based on standard NMR radio frequency (RF) shaped pulses, but, due
to the large number of degrees of freedom, a computer compiler was
built to systematically, numerically optimize the pulse sequences.
Using similar techniques, Vandersypen et al. demonstrated the simplest
instance of Shor's factoring algorithm on a seven-qubit register
\cite{Vandersypen:2001a}. At the same time, Cory's group at MIT
developed a different control approach based on the design of
numerically optimized high-fidelity pulses that refocussed all
unwanted interactions \cite{fortunato:7599}. More recently, both
approaches have been used to control a 13-qubit system
\cite{negrevergne:170501}.

The availability of long-lasting
coherence and high-fidelity control makes NMR an ideal test-bed for
quantum control techniques. In this paper we present two approaches to
NMR quantum control that we have used. The first approach applies
optimal-control theory to achieve high fidelity and robust control
over small spin systems. The second is a scalable technique, where a
well chosen approximation allows for imperfect control elements to be
placed in sequence while controlling the error buildup. We
show how the two methods can be combined to apply optimal control
methods to larger systems in a scalable manner.
The sequence of basic operations needed to implement an algorithm can
be described in the circuit model of quantum computing. In this model
the sequence of operations is given by a quantum circuit,
a time-ordered sequence of quantum gates to be applied to the
quantum systems. A typical quantum circuit has three main components:
initialization, implementation of the unitary evolution that realizes
the desired algorithm, and a final measurement to obtain the
algorithm's output. In this paper we focus on the unitary evolution.
We show how one can modulate the available control parameters to
implement a universal set of quantum gates and how one can combine the
gates to realize a complete quantum circuit.

The paper is divided into the following sections: In section
\ref{NMR_section} the relevant aspects of the system and principles of
control in an NMR QIP are explained; in section \ref{control_section} we
describe both control methods: an optimal control method based on an exact description of the system,
and an efficient approximation scheme based on subsystems; in section
\ref{merge_section}, we explain why they form a basis for a
flexible and general framework to design practical quantum control
methods. More details about the technical implementations of such
methods for liquid state NMR can be found in the appendices.

\section{NMR}
\label{NMR_section}
The basic principles of NMR are available in many textbooks
\cite{Ernst:1990a,Levitt:2001a}, and their use in quantum computing has
been reviewed in
\cite{Cory:2000fk,vandersypen:1037,Vandersypen:2001b,Baugh:2007a}. Here we discuss
only the relevant basics. In liquid-state NMR the qubits are the
spin-$\frac{1}{2}$ nuclei of a molecule. The number of
spin-$\frac{1}{2}$ nuclei in a molecule gives the number of qubits. A
sample with approximately $10^{20}$ identical molecules dissolved in a
liquid solvent is placed in a large static magnetic field that
provides the quantization axis and is conventionally taken to define
the z axis. Ideally all the molecules experience the same fields and
so undergo identical evolution. Since each molecule realizes a QIP,
this is an example of ensemble quantum computing.

The nuclei interact with the static field, which produces a Zeeman
splitting between the two energy levels (spin aligned or anti-aligned
with the static field), giving the Hamiltonian
\begin{equation}
\label{HZeeman}
\Ham_{Zeeman} = \gamma_iB_o(1+\delta)Z = \omega_iZ,
\end{equation}
where $\gamma_i$ is the gyromagnetic ratio of nucleus $i$, $B_o$ is
the strength of the magnetic field and $Z$ is the Pauli operator
$\sigma_z$ (similarly, we use $X$/$Y$ for $\sigma_x$/$\sigma_y$). We
always assume that the static magnetic field is along the z direction.
It is convenient to use a semi-classical picture of the spins
precessing about the magnetic field with an angular frequency
$\omega_i$.  The Larmor frequency is $\omega_i/2\pi$ and is on the
order of tens to hundreds of megahertz in today's superconducting magnets. In
the absence of symmetry of the molecule, each nuclear site in the
molecule experiences a different electronic environment. The electron
cloud partially shields the nuclei from the applied magnetic field so
that the local magnetic field experienced by each nucleus is slightly
different. This gives each nucleus a slightly different Larmor
frequency called the chemical shift, denoted by $\delta \ll 1$ in Eq.
\ref{HZeeman}. This separation in frequency space(of a few kilohertz) is
what allows different nuclei of the same nuclear species to be
distinguished and selectively addressed. Different nuclear species
have different gyromagnetic ratios, so their Larmor frequencies are
usually widely separated by tens or hundreds of megahertz in
frequency space. The contrast between the homonuclear and
heteronuclear frequency ranges leads to huge differences in the
timescales for homonuclear and heteronuclear control. It is only when
a purely heteronuclear system is considered that single-qubit rotations
are much faster (approximately three orders of magnitude)  than two-qubit coupling gates.

The nuclear spins interact with each other by two mechanisms: a direct
spin-spin coupling via the dipolar interaction, and an indirect
electron-mediated interaction, the J coupling. In the liquid state,
the rapid tumbling motion of the molecules averages out the dipolar
coupling between the nuclei, both within the same molecule and (to a
good approximation) between molecules \cite{Levitt:2001a}. This leaves
only the weaker and isotropic J coupling. The coupling
Hamiltonian between spins $i$ and $j$ is of an exchange form,
\begin{equation}
\Ham_{Jcoupling} = \frac{\pi}{2}J_{ij}\left(X_iX_j + Y_iY_j + Z_iZ_j\right).
\end{equation}
Since this coupling is mediated by the electrons binding the two
nuclei, there is no J coupling between different molecules. If the
difference in the resonant frequencies of the two nuclei involved is
much greater than the coupling strength, then a secular approximation
is valid. The off-diagonal terms can be ignored and the coupling is
reduced to the weak-coupling Ising form where the $X_iX_j$ and
$Y_iY_j$ terms are dropped:
\begin{equation}
\Ham_{Jcoupling_{weak}} = \frac{\pi}{2}J_{ij}Z_iZ_j.
\end{equation}

This approximation is always valid for heteronuclear systems and for
the majority of the homonuclear systems considered for a QIP. With this
approximation, all the terms in the internal Hamiltonian commute,
and tracking the system during periods of free evolution becomes
particularly straightforward. This can be used to simplify the
control schemes.

The internal Hamiltonian enables one-qubit rotations about the $Z$
axis and two-qubit controlled-$Z$ gates. However, for universal
control, one-qubit rotations about another axis are needed. These are
implemented by control-field Hamiltonians. By applying an RF field, we
can induce transitions between energy levels whose energy difference
is resonant with the field. In the rotating frame the
control field Hamiltonian can be described as,
\begin{equation}
\Ham_{control}  = \omega_{nut}(t)\left(\cos(\phi(t))X + \sin(\phi(t))Y\right).
\end{equation}
Up to the hardware limitations of the spectrometer, it is possible to
arbitrarily control the amplitude ($\omega_{nut}$, the nutation or
Rabi frequency) and phase ($\phi$). With a standard liquid-state probe,
nutation frequencies of up to 30 kHz are feasible. If the RF
frequency is on resonance with the spins' Larmor frequency, then in
the rotating frame the contribution to the Hamiltonian of the static
magnetic field along $Z$ vanishes and the RF field looks like a
constant field about an axis $\phi$ from the x-axis. Hence, the spins
precess about this axis at a frequency $\omega_{nut}$, and we can
induce rotations about any axis in the $X$-$Y$ plane in the rotating
frame.  In the pulsing frequency rotating frame, spins whose
transition frequencies are off-resonant have an additional $Z$
component in their Hamiltonian, and the effective rotation axis will be
the vector sum of the RF field and the off-resonant $Z$ field. If
the spins are far off-resonance then this rotation axis will be close
to the z axis and the spins will not be rotated into the plane.
However, the spins are not completely unaffected and pick up an
additional phase or z rotation known as the transient Bloch-Seigert
shift \cite{Emsley:1990a}, which must be accounted for (see section
\ref{compilercontrol}). The combination of the internal Hamiltonian and
the control fields gives universal control i.e. it is possible to
implement any unitary transformation on the system. The challenge is
to find the control fields as a function of time that drive the system
through the desired evolution.

Some subtle effects arise from the fact that NMR QIP's utilize an ensemble of quantum systems,  
rather than a single one, as is commonly
assumed in quantum computing:

\emph{State preparation.} In the liquid state the sample temperature
cannot vary much from room temperature. Even with the very strong
fields available, the Zeeman energy splitting is much smaller than the
available thermal energy and so the thermal state is highly mixed, with
only a very small ($\epsilon \sim10^{-5}$) bias towards the ground
state. The thermal state for a single spin is of the form
\begin{equation}
\rho_{thermal} = \frac{1}{2}\openone + \frac{\epsilon}{2} Z = \frac{1-\epsilon}{2}\openone + \epsilon\ketbra{0}{0}.
\end{equation} 
In NMR quantum information processing experiments, the evolution of the state is to a good
approximation unital, meaning that the identity component of the state
is preserved (the experiment time is much shorter than $T_1$). In addition, the identity component is not directly
observable because it has no effect on sample magnetization. For our
purposes, it is sufficient to observe the dynamics of the second
component of the state, called the deviation density matrix, which is
pure. This component is called the pseudo-pure state. The idea of and
method for preparing pseudo-pure states can be generalized to multiple
qubits:
\begin{equation}
\rho_{pp} =  \frac{1-\epsilon}{N} \openone + \epsilon \ket{\psi}\bra{\psi},
\end{equation}
where $\ket{\psi}$ describes a pure quantum state on $n$ qubits and
$N=2^n$ is the dimension of the Hilbert space. Unfortunately, the
standard preparation methods
\cite{DavidG.Cory03041997,NeilA.Gershenfeld01171997,PhysRevA.57.3348}
are inefficient and $\epsilon$ decays exponentially with the number of
qubits in a pseudo-pure state. Thus the signal quickly disappears in
the noise, and this limits the use of traditional NMR methods to
manipulate quantum information to less than approximately 15 qubits.
Furthermore, because the initial state is so highly mixed, the states
are provably separable, and a classical model, although somewhat
contrived, can describe the results of NMR experiments
\cite{PhysRevLett.88.167901}. Nevertheless, one can still use
pseudo-pure states to explore quantum control issues. It should be
emphatically stated that the efficiency of pseudo-pure states is not a
fundamental limitation to the scalability of an NMR QIP, and scalable,
efficient algorithms have been developed to create pure states with
NMR \cite{Boykin03192002,Schulman:1999a}. These algorithms are not yet
practical with current technology in liquid state NMR. However, they
are practical in other magnetic resonance systems where high
polarization is achievable, such as solid-state NMR
\cite{baugh:022305}, electron-spin resonance (ESR)
\cite{mehring:052303}. We expect the control techniques developed here
to readily transfer to these systems.

\emph{Measurement.} In standard implementations of a QIP, the ability
to make single-qubit projective measurements is assumed. However, in
the ensemble framework of NMR we can measure only the expectation
value of certain observables averaged over the ensemble of identical
systems. Fortunately, most quantum algorithms can be modified to fit
this framework \cite{NeilA.Gershenfeld01171997}. Furthermore, the
ensemble measurement is also sufficient to measure the fidelity of any
unitary transformation applied to an initial pseudo-pure state. This
makes it possible to benchmark control processes. There are some
subtle issues regarding the coarseness of the measurement, the size of
the system and how much the system is disturbed by the measurement
\cite{poulin:022102}; however, for  typical samples these can be
ignored.

Despite the problems due to the ensemble nature of the system, the
resulting Hamiltonian and control principles are applicable to many
different systems. Examples include other spin systems, such as
solid-state NMR \cite{baugh:022305}, or electron-nuclear systems in
organic crystals \cite{mehring:052303}, quantum dots
\cite{Koppens:2006uq} or nitrogen-vacancy centers in diamond
\cite{L.Childress10132006}, and even non-spin systems governed by
effective spin Hamiltonians such as superconducting qubits
\cite{Wendin:2005a}. The control principles developed should therefore
find application to these other, potentially more practically scalable
systems.

\section{Control methods}
\label{control_section}
Standard approaches to control in NMR spin systems are well known
\cite{vandersypen:1037} and have permeated other quantum information processing
technologies. In the present work we describe two different approaches
taken by our group to design and optimize pulse sequences, each
addressing an important issue. First, given an accurate model of our
system and apparatus, we can design high-fidelity control sequences by
simulation and optimization. Second, we make some well chosen
approximations to our model to design accurate and robust sequences
\emph{efficiently}.

\subsection{Optimal Control Theory}
By applying standard optimal control theory methods to the problem of
controlling quantum systems, high fidelity control sequences can be
found. The basic idea is to let some form of numerical optimization
search for the pulse shape of the control fields that exactly
implements the desired unitary gate: the pulse will internally refocus
all unwanted couplings and Bloch-Siegert shifts. Although we are left
with little physical intuition about the path the system is driven
through, very high fidelity control sequences are possible in a
variety of situations where simple shaped pulses do not work. Two
concerns with this powerful approach are its efficiency and
practicality. In principle, since universal control is available,
there exists a pulse shape that will implement any desired unitary.
However, finding it requires a simulation of the full system which
limits the optimal control approach to systems of less than
approximately 8 qubits  on a desktop computer and 10 qubits on larger clusters \cite{parallelGRAPE}. In addition, the pulse may require many hundreds if
not thousands of time steps that need to be optimized, giving a large
parameter space over which to optimize. A huge search space combined with
expensive function evaluations leads to a difficult optimization
problem.

Quantum optimal control is well studied in the context of driving
chemical reactions with shaped laser pulses. However, this addresses a
state transformation problem, whereas for quantum computing purposes,
it is necessary to solve the much more difficult problem of realizing
a unitary transformation that works for all input states. The total
Hamiltonian is the sum of the RF control Hamiltonians and the internal
Hamiltonian, \begin{equation}
\Ham_{tot}(t) = \Ham_{int} + \Ham_{rf}(t).
\end{equation}
As discussed in section \ref{NMR_section}, the internal Hamiltonian is
of the form $H_0 = \sum_i \omega_i Z_i + \sum_{i<j} J_{i,j} Z_i Z_j$.
This gives us Z rotations and controlled Z evolution, which is the
basis of the two qubit gates. The internal Hamiltonian is
time-independent and its parameters can be obtained with high
precision from standard spectroscopic techniques and fitting software.
The free parameters that can be used to control the evolution of the
system are in the RF~Hamiltonian: $H_{rf} = u_x\left(t\right)\sum_kX_k + u_y\left(t\right)\sum_kY_k$, where $k$ lists
the spins sensitive to the RF pulse, which is the spins of the same
nuclear species. The optimal control problem is to find the values of
the control fields $u_{x/y}\left(t\right)$ that ensure that the total
unitary evolution is as close as possible to the desired one. There
will be many control field sequences that give the correct evolution, but to
minimize decoherence the time-optimal solution is preferred. For
relatively simple systems with two or three qubits, and with a few
potentially unrealistic assumptions on the experimentally available
control, this can be solved analytically
\cite{carlini:042308,PhysRevA.63.032308}. However, for bigger systems
it is necessary to use a numerical optimization procedure to find
control sequences. The
algorithms start with a random guess for the control fields, which are
discretized at a sufficiently high resolution in time. The system is
then simulated and the unitary $U_{sim}$ is obtained. This
is then compared with the goal unitary $U_{goal}$ through a well
chosen fitness function. Since global phases are irrelevant, a
suitable choice that gives freedom in the global phase is the 
Hilbert-Schmidt (HS) fidelity, defined by 
\begin{equation}
\label{tracefidelity}
\Phi = \left| tr\left(U_{goal}^\dagger U_{sim}\right)\right|^2/N^2.
\end{equation}
This fidelity measure can be seen as an imperfect motion reversal and
is linearly related to the average fidelity, defined as the squared
state overlap between the ideal output state and the simulated output
state averaged (Haar measure) over all input states \cite{Nielsen:2002a}. Given this fitness function, any numerical optimization
scheme can be used to search for the highest-fidelity pulse.

The first work on numerically optimizing RF control for an NMR QIP was
performed by Cory's group at MIT \cite{fortunato:7599}. Their approach
was to simplify both the search space and the function evaluations by
using pulses consisting of blocks of constant amplitude, phase and
frequency. Approximately 10 to 30 of these blocks were combined into a
composite pulse called a ``strongly modulating pulse''. Thus there
were only 10 to 30 time steps to evaluate and compose in order to compute
the overall unitary. The parameter space of the composite pulse was
then searched with a simplex search algorithm. This method
successfully found high fidelity pulses that were experimentally
implemented in a variety of quantum information processing demonstrations \cite{hodges:042320}. However, the search method is
not optimal and has difficulty finding high fidelity pulses for more
than four qubits. In addition, the pulses have sharp discontinuities
between the blocks that lead to transient effects in pulse generation
and in the NMR probe resonant circuit and suboptimal experimental
implementation.

A substantial improvement was made by applying standard optimal
control theory to the problem. This resulted in the GRadient Ascent
Pulse Engineering or GRAPE algorithm of Khaneja et al.
\cite{Khaneja:2005fk,schulte-herbruggen:042331}.   If a pulse is digitized the control fields become a sequence $u_i(j)$ where $j = [1...M]$  denotes the $j^{th}$ timestep.  The unitary for each timestep is,
\begin{equation}
U_j = exp\left[-i\Delta t\left(\Ham_{int} + \sum_iu_i\left(j\right)\Ham_i\right)\right].
\end{equation}
Where $\Ham_x = \sum_kX_k$ for example and $\Delta t$ is the length of the timestep.   To first order, the derivative of the unitary propagator with respect to the control fields can be evaluated without finite differencing or another matrix exponential as,
\begin{equation}
\label{approxgradient}
\frac{\delta U_j}{\delta u_i\left(j\right)} \approx -i\Delta t \Ham_i U_j
\end{equation}
where we require $|\Delta t \Ham| \ll 1$ for the approximate derivative to be accurate.  The total unitary for the pulse can be calculated as the product of each timestep unitary, 
\begin{equation}
U_{tot} = U_MU_{M-1}U_{M-2}....U_3U_2U_1.
\end{equation}
The gradient of the fitness function Eq. \ref{tracefidelity} can now be explicitly calculated as
\begin{multline}
\label{totalgradient}
\frac{\delta\Phi}{\delta u_i\left(j\right)} =  \frac{1}{N^2} \times \\ \bigg[ tr\left(\left(U_{j+1}^\dagger...U_M^\dagger U_{goal}\right)^\dagger \frac{\delta U_j}{\delta u_i\left(j\right)}U_{j-1}... U_1\right)  + c.c. \bigg]
\end{multline}
Thus by storing the forwards and backward propagation of the unitary and substituting Eq. \ref{approxgradient} into Eq. \ref{totalgradient}, gradient information about the fitness function can be obtained without finite differencing or recalculation of the entire propagator.  The gradient leads to a much more efficient search determining the direction in which the control parameters should be moved to improve the fitness function.  This information can be used by simple steepest-ascent hill-climbing algorithm to optimize the pulse.

 An important advantage of the GRAPE
method is that the more efficient search method allows the pulse to be
defined with many more points so that smooth ``spectrometer friendly"
pulses can be obtained, as explained in the appendix. Our code is based
on the published algorithm \cite{Khaneja:2005fk} with some
modifications also discussed in the appendix. It finds high fidelity
(above $ 99.75\ \%$ HS fidelity including experimental inhomogeneities in
RF and static fields) in a number of systems with up to seven
qubits. The optimization method finds only local maximum in
the search space so that better pulses may exist, but the local maxima
found have proven to be sufficient for high fidelity control.

\begin{figure*}[htbp]
\includegraphics[scale=0.55]{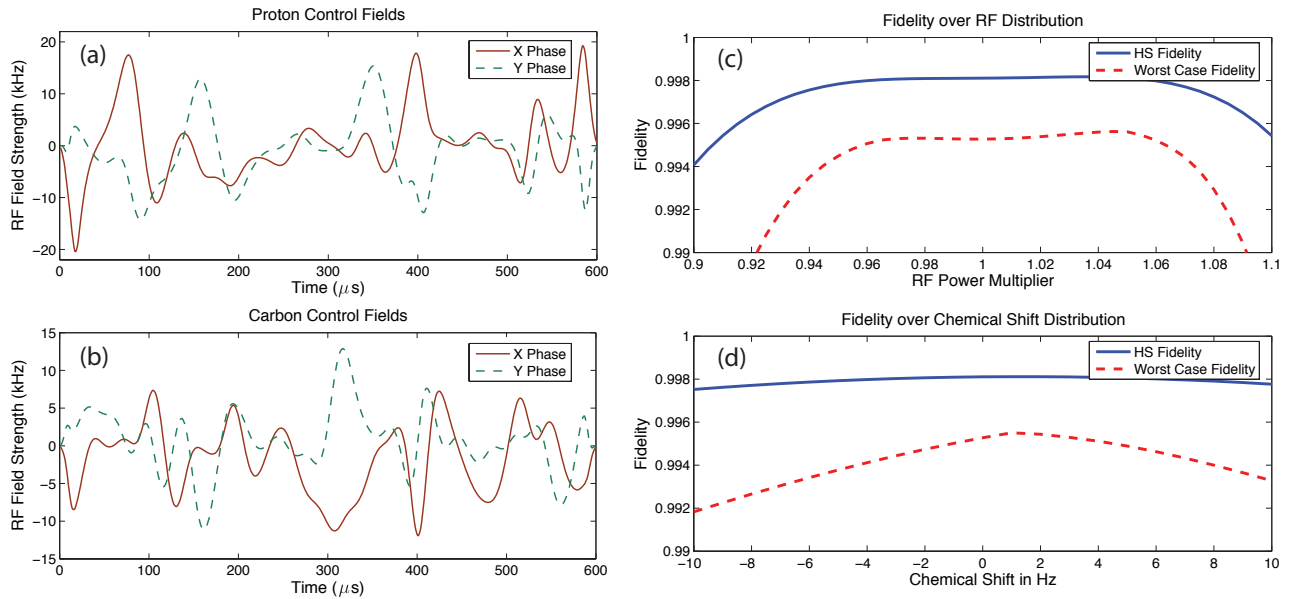}
\caption{\label{GRAPEpulse} Plot showing the quadrature components of
the proton (a) and carbon (b) control fields for a GRAPE pulse that
implements a 90 degree rotation on $H_1$ in the crotonic acid
molecule (see Appendix \ref{crotonic_appendix} for details of molecule). This pulse is one of the more difficult to find one-qubit
gates on the crotonic acid molecule, as $H_1$ and $H_2$ are only 788
Hz apart on a 700 MHz spectrometer. The pulse was found by using the
subsystem technique described in Appendix \ref{GRAPE_appendix}. The
pulse shape is smoothly varying and goes to zero at the start and end
points so as to avoid any transient effects in implementation. The
total duration of the pulse is only $600\ \mu s$, compared with the
over $3\ ms$ a standard soft pulse would take. This pulse is not time
optimal, but shorter pulses tend to require unfeasibly high power. The
right-hand-side plots show the robustness of the pulse in the presence
of two common experimental inhomogeneities. Both the HS
(\ref{tracefidelity}) and the worst case fidelity (the overlap of the
output state from the simulated unitary and the output from the ideal
unitary minimized over all input states) are shown. The fidelity of
the pulse shows an almost flat profile with respect to variation in
both the RF field strength and static field. Thus, the performance of
the pulse is unaffected by inhomogeneities or miscalibration. This
robustness also obviates the need to find new control pulses due to
small changes in the internal Hamiltonian over time.}
\end{figure*}

The numerically optimized control sequences drive the system through a
complicated and nonintuitive path. As a result it is possible that
small errors in the model of the system and apparatus could lead to a
much lower fidelity pulse. Fortunately, the optimization method can be
modified so that the pulses are robust in the presence of static
inhomogeneities in the system and the control Hamiltonian. For
example, due to mismatch of magnetic susceptibilities of the sample
and sample tube and imperfect shimming, the static magnetic field
varies across the sample. We can demand that the control sequence
implement the same desired unitary for a range of static magnetic
fields, so that the pulse is robust given the corresponding variations
in the Larmor frequencies. Another practical example is that the
amplitude of the control fields may not be perfectly calibrated and
again vary across the sample because of coil geometry, so we can also
demand that our pulse implement the desired unitary for a range of
control powers. Robustness with respect to both these effects for a particular
seven-qubit pulse is shown in Figure \ref{GRAPEpulse}.

\subsection{Efficient pre-simulation and optimization}
\label{compilercontrol}
Because the classical cost of simulating the full quantum dynamics
grows exponentially with the system size, it is not possible to apply
optimal control in larger systems. Some well-chosen simplifications
and approximations must be made to the model. The control sequence is
constructed from simple predefined but imperfect building blocks (for
example standard frequency selective pulses). The blocks are then
systematically put together to form a pulse sequence ensuring that the
errors in the building blocks do not build up excessively as the
sequence progresses. It is possible to efficiently design such
sequences only if we judiciously simplify the model and take into
account only the largest and first order errors in the blocks.

\begin{figure}[htbp]
\includegraphics[scale=0.475]{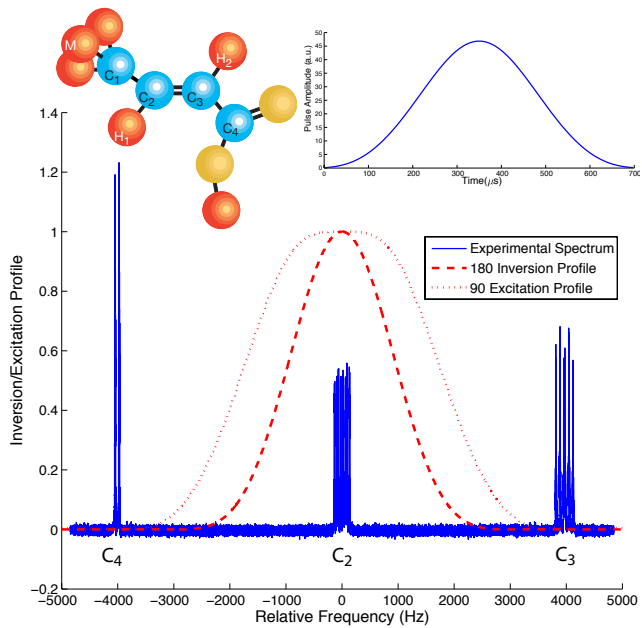}
\caption{\label{simplepulse} A portion of a single-scan carbon thermal
spectrum of the fully labeled molecule trans-crotonic acid
\protect{\cite{Knill:2000a}} at 16.4T, showing three qubits,
$C_2$,$C_3$ and $C_4$. See Appendix \protect{\ref{crotonic_appendix}}
for details of the molecule. By modulating the amplitude of a pulse
sent at a frequency resonant with the chemical shift of $C_2$ we can
selectively rotate $C_2$. Shown in the inset is a truncated inverse
secant pulse of $700\ \mu s$. Overlain on the spectrum is the
excitation profile of a 90 degree pulse and the inversion profile of a
180 degree pulse of this length. The profiles are expressed
as a fraction of the goal tipping angles. It is clear that the pulse affects
$C_2$ but does not significantly excite $C_3$ or $C_4$. As discussed
in the text, $C_3$ and $C_4$ experience a transient Bloch-Siegert
effect, which must be accounted for in a quantum computing experiment.
}
\end{figure}

Typical building blocks are one-qubit unitaries that involve
selectively rotating one spin. If the spins have distinct resonant
frequencies, this corresponds to frequency selective pulses. The
problem of obtaining such pulses has of course had much attention in
the long history of NMR. The most successful approach has been to use
shaped pulses, and a huge variety of increasingly complicated shaped
pulses have been developed with various bandwidths, excitation
profiles (tipping angles as a function of chemical shift) and
robustness to experimental imperfections \cite{Freeman:1998a}. In the
linear regime the excitation profile in frequency space of a pulse is
related to its Fourier transform
\footnote{The linear regime applies only to small angle rotations but
it still gives good intuition for larger rotation angles.}. As one
would expect, the longer the pulse, the more selective it is in
frequency space. Furthermore, one can tailor the excitation profile by
shaping the pulse. For example, a Gaussian shaped pulse has a Gaussian
shaped excitation profile. Thus, given the internal Hamiltonian of the
molecule, it is straightforward to design a set of pulses for single
spin rotations. An example of a pulse and its excitation profile is
shown in Figure \ref{simplepulse}.

For many NMR spectroscopy experiments, it is sufficient to
consider only the excitation profile, but for quantum computing purposes
we need to accurately keep track of all the effects of the pulse. In
particular there are off-resonant and coupling effects. Although the
resonant frequency of a particular spin may be far outside the
excitation bandwidth of a selective pulse intended for another spin
(so that there is no rotation about an axis in the plane), there
still is a substantial phase evolution from the off-resonant or
transient Bloch-Siegert effect \cite{Emsley:1990a}. To first order,
the effect of an off-resonant pulse is to shift the resonant frequency
of the spins, inducing an extra phase rotation. For example, a spin
3 kHz off-resonant from a 1 ms pulse that performs a 180 degree rotation
on an on-resonance spin will experience an additional phase shift of
$\sim15$ degrees. This is non-negligible and must be accounted for.
Secondly, the couplings still evolve during the pulse (which may have 
a duration comparable to $1/2J$), and the coupling Hamiltonians do not
commute with the RF Hamiltonian. Particularly for long pulses on
spins that have strong couplings, there are substantial
deviations from the ideal action of the pulse. These can be partially
countered by self-refocussing pulses \cite{Freeman:1998a}, such as the
Hermite pulses (see Table \ref{subsystem_table}).

\begin{figure*}[htb]
\includegraphics[scale=0.4]{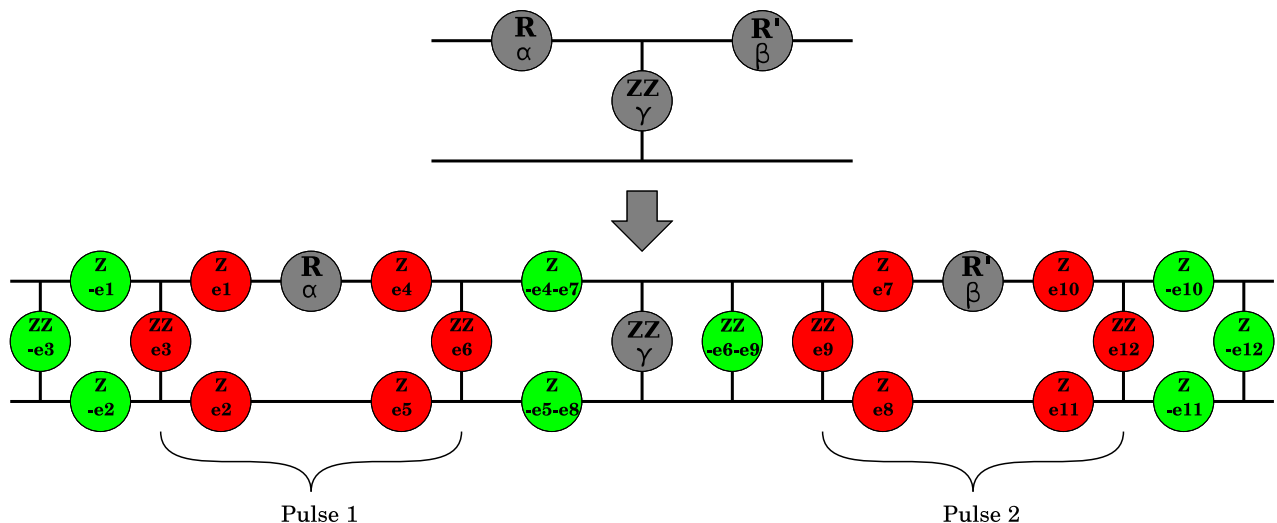}
\caption{\label{pulsedecomp} Sequence including pre- and
post-error terms and corrections. The circles denote rotations and couplings. The axis
and angle of the rotation are given one above the other. A rotation
around the ZZ axis by $\theta$ is the evolution $e^{-iZ_jZ_k\theta/2}$
for the target spins $j$ and $k$. Grey circles are for the intended
sequence. The red circles represent the pulse pre- and post-error terms,
while the green ones represent the evolutions that are implemented by the sequence compiler in order
to cancel the pulse error terms. The ZZ-corrections are implemented by
adjusting the free-evolution periods between pulses and the
Z-corrections are performed with phase tracking calculations. }
\end{figure*}

Both the off-resonant and the coupling effects can be accounted
for through a decomposition scheme where the imperfections are
unravelled from the simulated pulse and represented as phase and
coupling errors before and after the ideal pulse. A similar method
with symmetric negative time evolution was presented in Ref.
\cite{Vandersypen:2001b}. The method described here is more accurate
and general. The simulated pulse is modeled by the following
decomposition:
\begin{align}
\label{decomp}
U_{sim} & =   e^{-i (H_{rf} + \sum_i(\omega_i \textbf{Z}_i)  +\sum_{i<j}(J_{ij}\textbf{Z}_i\textbf{Z}_j))\Delta t} \notag\\
      &\simeq  \prod_i e^{-i\alpha^{post}_i \textbf{Z}_i } \prod_{i<j}e^{-i(\beta^{post}_{ij}
    \textbf{Z}_i \textbf{Z}_j)} U_{ideal} \times \notag \\
    & \qquad\qquad\qquad \prod_i e^{-i\alpha^{pre}_i \textbf{Z}_i } \prod_{i<j}e^{-i(\beta^{pre}_{ij} \textbf{Z}_i \textbf{Z}_j)}
 \end{align} 
This decomposition resolves the simulated pulse into the ideal unitary
operator sandwiched on either side by Z rotations (to account for the
Bloch-Seigert shift and the chemical shift evolution during the pulse)
and ZZ couplings to account for the couplings that occurred during the
pulse (see Fig. \ref{pulsedecomp}).  This model for the pre- and post-error terms does not work for arbitrary pulses, but in the relevant case of spin selective pulses, it will capture most of the first order dynamics.  All couplings that do not involve the target qubit are trivial to extract and can be perfectly reversed.  For couplings involving the target qubit, the control fields modulate the coupling necessitating a numerical optimization of the error terms.  For 90 degree pulses ZZ pre- and post-error terms can represent coupling evolution during the pulse.  However, for 180 degree pulses, because the ideal pulse should refocus the couplings, ZZ error terms are not sufficient, and self-refocussing pulses are desirable. Since the pre- and post-error terms do not commute with the control Hamiltonian, the decomposition faithfully
represents the true dynamics only when $\left|\Delta tJ\right| \ll 1$ where $\Delta t$ is the length of the pulse and $J$ is the strongest coupling. 

 The optimal parameters for the Z and ZZ ``error'' terms can be determined by use of an efficient procedure. A set of single-spin and pairs-of-spins simulations is sufficient to
capture off-resonant effects and some first-order coupling
contributions, since the Hamiltonian contains only one- and two-body
terms, and all the pre- and post-error terms commute with each other. The simulations are of a fixed one- or two-qubit size, hence there are
only $n+ \frac{n(n-1)}{2}$ simulations to perform. From these
simulations, the optimal pre- and post-error terms are determined by use of
a numerical search. See Appendix \ref{pulsecompiler_appendix} for a
detailed description of how to obtain the error terms. The key points
are that an efficient approximate description of the pulse is possible,
and that the deviations from the ideal pulse in the decomposition can
be corrected later as part of a larger pulse sequence by means of
periods of free evolution and individual rotating frames adjustments
(as shown in Fig. \ref{pulsedecomp}). An example decomposition of the
pulse implementing a 90 degree rotation on $C_2$ (shown in Figure
\ref{simplepulse}) is given in Table \ref{pulsedecomp_numerical}. The
representation of the pulse as Eq. \ref{decomp} has an average
fidelity of 99.96\ \% with respect to a simulation of the full system
dynamics.

\begin{table}[htdp]
\begin{ruledtabular}
\begin{tabular}{| d || d | d | d | d | d | d | d |}
\hline  &  M  &  H_1  &  H_2  &  C_1  &  C_2  &  C_3  &  C_4  \\
\hline\hline     M  &   0.0 & 0.41 & -0.04 & 7.95 & -0.41 & 0.41 & -0.06 \\
\hline H_1 & 0.41 & 0.0 & 0.98 & 0.18 & 11.32 & -0.11 & 0.41 \\
\hline H_2 & -0.04 & 0.98 & 0.0 & 0.34 & 0.0 & 10.07 & 0.24 \\
\hline C_1 & 7.95 & 0.18 & 0.34 & 1.72 & 2.98 & 0.09 & 0.44 \\
\hline C_2 & -0.41 & 11.32 & 0.0 & 3.00 & 0.0 & 4.95 & 0.09 \\
\hline C_3 & 0.41 & -0.11 & 10.07 & 0.09 & 4.95 & 7.12 & 4.40 \\
\hline C_4 & -0.06 & 0.41 & 0.24 & 0.44 & 0.09 & 4.40 & -6.92 \\
\hline 
\end{tabular}
\end{ruledtabular} 
\caption{\label{pulsedecomp_numerical} Table showing the decomposition
of the single spin and pairwise simulations for the pulse performing a
90 degree rotation on $C_2$ in the crotonic acid molecule. The
diagonal gives the Bloch-Seigert shift in degrees with respect to each
nucleus' rotating frame, although in practice
everything is calculated with respect to a single reference frame. The
off-diagonal elements give the pre- (above diagonal) and post- (below
diagonal) ZZ error terms in degrees. Because the pulse is symmetrical,
so are the error terms, but the method is general enough to
handle arbitrary pulses. }
\end{table}

These pulse representations give us the building blocks for one-qubit
gates. Two-qubit gates are achieved by periods of free evolution
during which the coupling terms in the Hamiltonian evolve. If a
coupling term is allowed to evolve for a time $1/2J$, then a
controlled-$Z$ gate (up to one-qubit $Z$ rotations) is implemented
between the coupled spins. All undesired couplings must be refocussed
with 180 degree pulses. Putting the pulses and delays together to form
a pulse sequence that implements a desired algorithm has been
automated through the use of a sequence compiler.

Several further simplifications are used. Individual rotating frames
are defined for each spin. Just as the rotating frame for a single
spin eliminates the Zeeman term in the Hamiltonian, individual
rotating frames eliminate all chemical shift terms from the internal
Hamiltonian. This could be implemented by a transmitter dedicated to
each spin and rotating at that spin's chemical shift, but this is
experimentally prohibitive. The phase evolution caused by the chemical
shift term in the Hamiltonian is simply a Z rotation, and this
evolution can be tracked with respect to some fixed reference frame.
Similarly the evolution of the transmitter at its frequency can be
tracked, and when a pulse is required, the pulse phase is adjusted to
obtain the correct phase in the spin's rotating frame.  Similarly the
observation phase must also be calculated and adjusted. This tracking greatly
simplifies the pulse sequences by eliminating the need to refocus the
chemical shift evolution. It also allows Z rotations to be implemented
instantaneously and with high precision through a frame change -- that
is by simply modifying the phases of subsequent pulses and potentially
the observation phase. Thus it is worth transforming the quantum
circuits to use Z rotations preferentially. A drawback of individual
reference frames is that it is not always possible for the pulse to
have the correct phase in two different reference frames, so global
pulses that nontrivially affect multiple spins are not always
feasible, and hence they must be decomposed into a sequence of simpler pulses.

The coupling gates are implemented with delay periods to allow the
desired couplings to evolve. Unfortunately, the undesired couplings
also evolve. The unwanted couplings can be refocussed with
a series of 180 degree pulses by means of standard, efficient
algorithms \cite{Jones:1999lr,PhysRevA.61.042310}. However,
it is not necessary to refocus all unwanted couplings during every
evolution period. It is sufficient to track the evolution, and refocus
it only when needed \cite{bowdrey:032315}. Single qubit gates and the
Z and ZZ error terms in their representation commute with the terms
of the internal Hamiltonian, except those involving the target nucleus.
Therefore, we can let couplings evolve through many gates and 
need to arrive at the goal coupling only when a noncommuting pulse affects
either member of the coupling pair. The refocussing scheme can be made
more efficient by the use of ``virtual 180s". Conventionally, every
time a refocussing pulse is used, a second refocussing pulse must
be applied at the end of the period to cancel the first one and ensure
that all the spins return to the their initial state, so that the
computation is unaffected by the refocussing. While the first
refocussing pulse is needed to physically refocus the coupling, the
second one can be made virtual and not physically applied. The virtual
180 can be implemented by pushing it forwards or backwards (the
virtual 180 can be created before or after the physically applied
refocussing pulse) through the pulse sequence until it can be merged
with another pulse. If it is another refocussing pulse, then it can
cancel with the virtual 180 created at that refocussing event. Or, if
it is a computational pulse, the virtual 180 can be absorbed by
modifying the phase of the pulse and introducing a frame change (see
Appendix \ref{pulsecompiler_appendix} for details).

All these techniques are combined into an efficient pulse compiler.
The pulse compiler loads the information about the internal Hamiltonian,
the effective pulse decompositions (Eq. \ref{decomp}) and the desired
pulse sequence with a suitable refocussing scheme. It can then
optimize the delays between the pulses, the timing of the refocussing
pulses and the phases of the pulses to best implement the desired
sequence. At any point in the sequence, phase or coupling error terms from
the decomposition are compensated for by use of the phase tracking
calculations or are absorbed in the refocusing scheme (See Figure
\ref{pulsedecomp}). This allows for long pulse sequences to be
constructed while avoiding excessive error accumulation. For further
details of the sequence compiler, see Appendix
\ref{sequencecompiler_appendix}.

The sequence compiler can optimize only the delays between events and
it must be given a suitable sequence of refocussing pulses to start
with. Designing an exact refocusing scheme may require many refocusing
pulses, and each refocussing pulse takes a finite duration and
introduces its own errors. In some cases, the theoretical control
accuracy gained by bringing the calculated $ZZ$ coupling evolution
closer to the goal is lost due to decoherence and pulse imperfections.
There is therefore a trade-off between the theoretical accuracy of
the control scheme and its duration and number of refocussing pulses.
In practice this entails designing a good refocussing scheme and
optimizing it. The total error from unrefocussed $ZZ$ couplings can be
calculated and determined whether it is acceptably low. If not, additional
refocussing pulses are added to correct the errors and improve the
optimization. In addition, a penalty function prevents the
optimization from using excessive time for refocussing schemes.

\section{Merging both methods}
\label{merge_section}

Each method described above has its strengths and limitations. Optimal
control theory can give robust and time-optimal control sequences for
strongly coupled systems, where conventional pulse design fails.
However, the method is intrinsically unscalable and is limited in the
number of spins it can handle in practice. The number of qubits
currently available in NMR already pushes these methods to their
limit. The pre- and post-error method with pairwise
simulations provides a scalable, efficient solution to the design of
control sequences but is limited by the properties of the pre-defined
pulses, particularly the fidelity of the error-term representation. On
long pulses targeting spins with strong couplings, the decomposition
of Eq.~\ref{decomp} may fail to give a high fidelity representation of
the pulse, and better pulse engineering is needed. Here we show how
the two methods can be combined to allow optimal control techniques to
be applied to larger systems.  A more computer science approach to scaling up optimal control techniques also utilizing subsystems has also been considered in Ref. \cite{Glaser:2007a}.

The idea is to consider only a subsystem of the QIP's qubits in
designing the optimal control pulse. In NMR, particularly relevant
subsystems consist of the spins of the same nuclear type, for example,
all protons. A pulse that is designed on the subsystem without
consideration for the other spins does not implement the desired
unitary on the whole system. To determine the effect of the pulse
on the entire system, pair-wise simulations between the subsystem and
the other spins are performed. These simulations capture and track
both the evolution of the internal Hamiltonian on the other spins, and
deviations due to couplings between the subsystems.  The pulse
on the full system can then be represented in the same manner as
described above by adding pre- and post-error terms, which can be
accounted for during the optimization of the refocusing scheme and the
phase tracking calculations as part of a larger sequence.

There is no guarantee that the pulse decomposition with the error
terms will give a high fidelity representation. The optimal control
pulse drives the subsystem through some complicated trajectory, and it
may not be possible to extract the effect of the couplings to other
systems as simple ZZ errors before and after the ideal gate. In
general this decomposition works well for short one-qubit unitaries
but breaks down for two-qubit gates taking more time. However, whereas
the pulse is optimized considering only subsystem 1, it can be made
robust to the effect of couplings to other subsystems by incoherently
averaging over the states of the other subsystems as described in Ref.
\cite{weinstein:6117}. This is equivalent to making the pulse robust
against Larmor frequency variations. Thus, averaging over all possible
states is not necessary, and making the pulse robust to frequency
shifts from the sum of the J couplings is sufficient. This will make
the pulse more difficult to find, but the obtained pulse will have a
higher fidelity representation on the full system.

\begin{figure*}[htbp]
\includegraphics[scale=0.5]{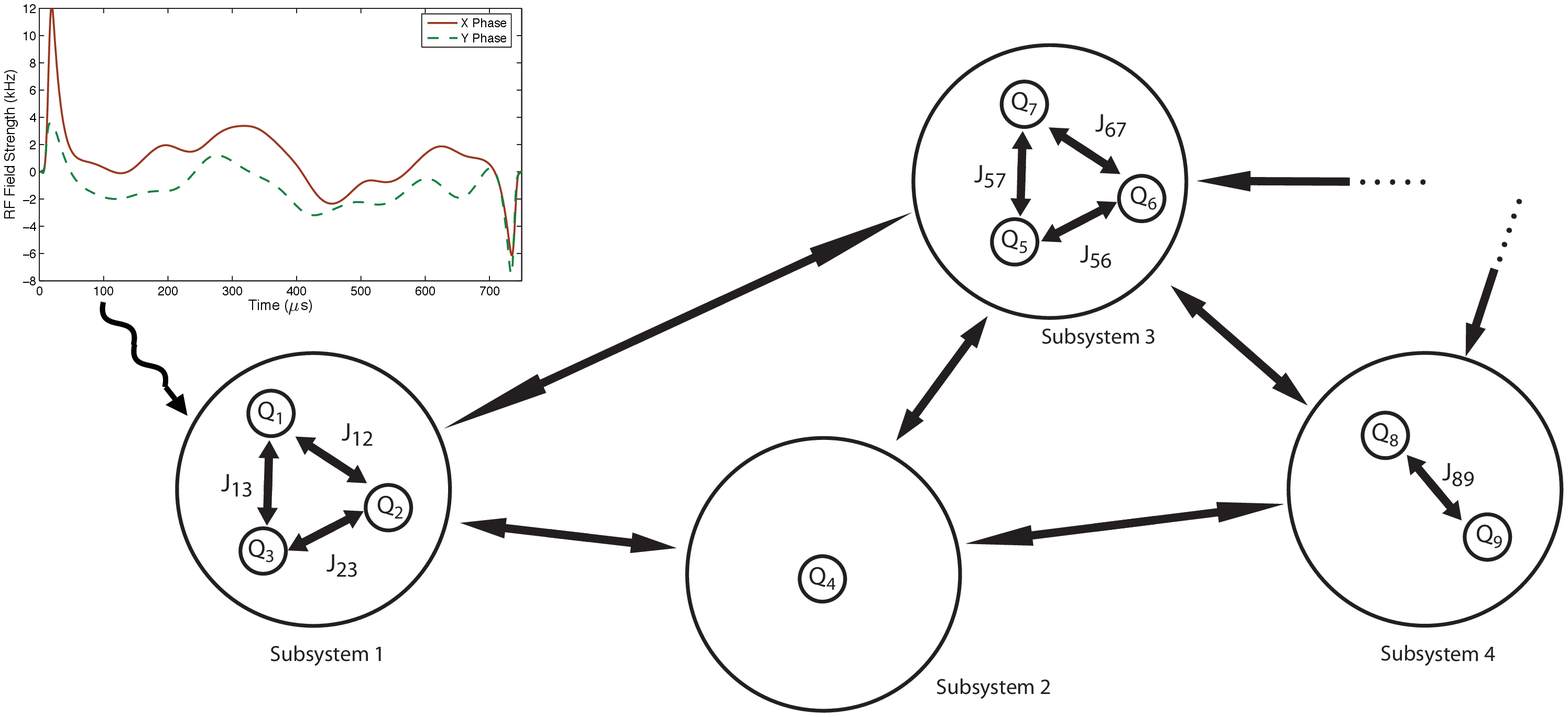}
\caption{\label{subsystems} Example of a block architecture: For
clarity, couplings between qubits of blocks have been contracted into
single arrows. It is assumed that the longer range couplings, such as
those between subsystem 1 and 4, vanish.  For example, in crotonic acid, subystem 1 could be the proton spins, so that $Q_1,Q_2,Q_3$ are $M,H_1,H_2$,  and subystem 3 the carbon spins and there would be no further subsystems.  In the inset an example of a sub-system optimal control pulse, in this case the pulse implements a 180 degree rotation on the methyl
qubit of the crotonic acid molecule, is found considering only
subsystem 1. The effects of the other subsystems are then
characterized using a set of ``subsystem'' pairwise simulations. As
previously, such simulations can be used to compute optimal pre- and
post-error terms that enable control of the couplings
by optimization of the refocusing scheme and the phase
tracking calculations described in Section \ref{compilercontrol}. 
}
\end{figure*}

To illustrate the basic ideas, consider a register of qubits organized
into subsystem blocks, as illustrated in Figure \ref{subsystems}. Take
for example the first subsystem of Fig. \ref{subsystems} consisting of
three proton spins, where the other subsystems consist of other
nuclear types. It may be that to effect a one-qubit rotation on, say,
qubit $Q_1$ by a simple method requires a very long selective pulse,
because the chemical shift difference to the nearest spin is small.
Then, the effect of the couplings cannot be taken into account by only $Z$  and $ZZ$ pre-
and post-error terms, because $\Delta t J$ is too large. However, one
can find an optimal control pulse that considers only subsystem 1 and
that refocusses the intra-subsystem couplings while implementing the
ideal gate. Perturbations due to inter-subsystem couplings and the internal evolution of the other subsystems during the pulse can then be taken into account by pre- and post-error terms.

The utility of such a scheme can be demonstrated on the crotonic acid
molecule (see Appendix \ref{crotonic_appendix} for details of  the molecule), where the natural subsystems are the three proton qubits and
and the four carbon qubits. Because of their refocussing properties,
180 degree rotations are more difficult to find and to represent.
However 180 degree rotations can easily be found for all the spins
by use of this subsystem technique and then represented through subsystem
pair-wise simulations. For example a pulse on methyl is found by 
considering only the proton subsystem, and then, to determine the
dynamics on the full system, simulations are performed on individual and
pairs of the five subystems, $\{M,H_1,H_2\}$, $\{C_1\}$; $\{C_2\}$;
$\{C_3\}$; and $\{C_4\}$. The results are summarized in Table
\ref{subsystem_table} and compared with the pulse representations of
standard pulse designs. The crotonic acid molecule is not ideally
suited to this approach because of the large couplings between
subystems; nevertheless, the subystem GRAPE pulses have similar or
better fidelities than standard pulses of a similar or longer length. In particular
the GRAPE pulses have consistently high fidelities even where other
pulses, such as selective pulses on $H_1$ and $H_2$, break down.
Furthermore, since the clock of decoherence is always ticking, shorter
pulses usually perform better.

\begin{table}[htdp]
\begin{ruledtabular}
\begin{tabular}{| c || c | c | c | c | c | c |}
\hline
& \multicolumn{2}{c|}{isech} & \multicolumn{2}{c|}{Hermite180} & \multicolumn{2}{c|}{GRAPE} \\
\hline\hline
& Length  & Fidelity &  Length & Fidelity &  Length & Fidelity \\
\hline
$M$ & 896 $\mu$m & 99.50 & 2m & 99.96 & 750 $\mu$m & 99.92 \\
\hline
$H_1$ & 3.3m & 91.71 & 7m & 97.71 & 3m & 99.79 \\
\hline
$H_2$ & 3.3m & 97.42 & 7m & 97.42 & 3m & 99.79 \\
\hline
$C_1$ & 128 $\mu$m & 99.92 & 300 $\mu$m & 99.97 & 60 $\mu$m & 99.95 \\
\hline
$C_2$ & 700 $\mu$m & 99.45 & 1.4m & 99.93 & 700 $\mu$m & 99.92 \\
\hline
$C_3$ & 700 $\mu$m & 99.39 & 1.4m & 99.94 & 700 $\mu$m & 99.85 \\
\hline
$C_4$ & 700 $\mu$m & 99.86 & 1.4m & 99.96 & 700 $\mu$m & 99.88 \\
\hline
\end{tabular}
\end{ruledtabular} 
\caption{\label{subsystem_table} Table comparing the performance of
optimal control pulses constructed and corrected by using the subsystem
approach (last column) to more conventional shaped pulses (first two
columns). All pulses implement one-qubit 180 degree rotations (see
text). The crotonic acid molecule is not ideally suited to the
subystem approach because of the large couplings between
subystems. This required all the pulses except for that on $C_1$ to
be made robust to the other subystems with incoherent averaging over
chemical shifts of the J coupling strength in order to obtain high
fidelities. }
\end{table}

Finding an optimal control pulse on the entire system such as in
Figure \ref{GRAPEpulse} may give shorter, higher fidelity pulses
than the subsystem with the error term strategy described above. This is
because the optimal control method is able to exploit more control
handles. However, optimization over the full system is too difficult
for large systems, whereas the methods used to find the pulses in
Table \ref{subsystem_table} are scalable if the subsystems used are
kept small. The combination of optimal control and use of pre- and
post-error terms is therefore well suited for designing control
sequence for the class of QIPs made of subsystems with strong internal
couplings but weaker coupling between subystems.
 
\section{Conclusion}
Liquid state NMR has led to the development of numerical methods to
find accurate pulses to control simple systems robust against errors
from perturbations in fields, amplitudes and other spins. Together
with methods for accounting for or refocussing phases and couplings,
we have a basis to \emph{efficiently} design \emph{robust} control
sequences for QIPs made of weakly coupled blocks of qubits. Within
blocks, faster dynamics are handled by pulses obtained by optimal
control methods. Pre- and post-error term analysis can be performed on
such pulses to evaluate perturbations due to the other blocks. On can
then combine such pulses to efficiently construct a sequence to
implement the desired algorithm. Phase tracking and numerical
optimization of the refocussing scheme allows us to correct pulse
errors, thus avoiding error accumulation, and to implement gates
between different subystems. The combined model of control presented
here can be applied to other Ising type models. A potential example is
in superconducting qubits, where the Hamiltonian can be written
in a pseudo-spin form that is similar to the liquid state NMR
Hamiltonian \cite{Wendin:2005a}. It is conceivable that the control
fields would affect only a small number of qubits at a time, thus
ensuring the presence of natural subystems to use for optimal
control.

\begin{acknowledgments}
We thank K. Rose for his assistance in coding the
sequence compiler, M. Silva and D. Kribbs for the algorithm to
deterministically compute the worst case fidelity, and M. Ditty for
technical assistance with the spectrometer. We also greatly appreciate
the helpful discussions with D. Cory and the editorial help
of Bryan Eastin and Scott Glancy. This work was funded by ARDA,
ARO and NSERC. Contributions to this work by NIST, an
agency of the US government, are not subject to copyright laws.
\end{acknowledgments}


\appendix

\section{Crotonic acid}
\label{crotonic_appendix}

Crotonic acid offers a total of seven qubits: two protons and the
methyl group (red) and four $^{13}C$-labelled carbons (blue). The
methyl group consists of three magnetically equivalent protons that
form spin-$\frac{3}{2}$ and spin-$\frac{1}{2}$ subspaces. With an
appropriate pulse sequence we can select the spin-$\frac{1}{2}$
subspace and treat it as a qubit. The internal Hamiltonian of the
system is shown below.

\begin{figure}[htb]
\includegraphics[scale=0.75]{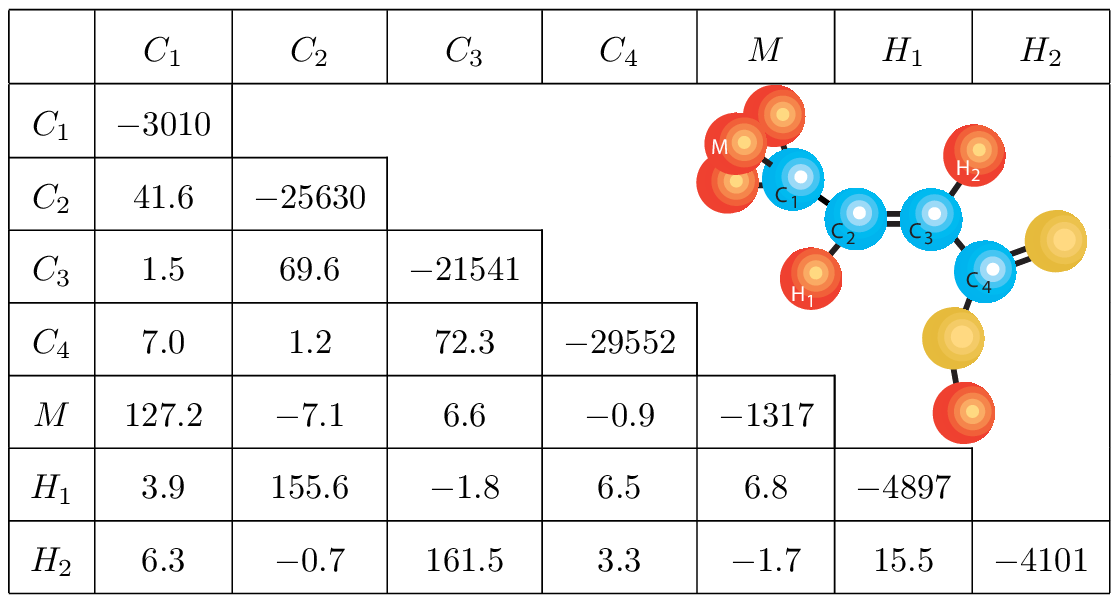}
\caption{\label{croto_table}Crotonic acid internal Hamiltonian
parameters as determined by spectral fitting in a 700 MHz spectrometer.
Diagonal elements give chemical shifts (in hertz) with respect to the base
frequency for proton and carbon transmitters. The off-diagonal
elements give J coupling terms also in hertz. }
\end{figure}

\section{GRAPE Pulse finder:}
\label{GRAPE_appendix}

The GRAPE code developed at IQC is closely based on the search
algorithm presented in Ref. \cite{Khaneja:2005fk}. Here we emphasize
only the specific modifications we have implemented that help to
achieve high experimental fidelity. The parameters for the initial random guess have a significant impact
on the success of the search and the type of pulse found. For good
experimental implementation, smoothly varying low power pulses are
best. Therefore the algorithm is initialized with a smooth low-power
guess. For example, in a 250 point pulse every 25th point is random
with a maximum range of $\pm 5\ \%$ of the maximum power available, and
cubic splines are used to interpolate the rest of the points. For
finding unitary gates, the fitness function is the HS fidelity defined
in Eq \ref{tracefidelity}. For good experimental fidelity the pulse is
made robust to variation in RF field strength and/or calibration
errors by taking an average of the fitness function defined over a
range of RF field strengths. A 3 point distribution at RF power
multipliers of 0.95, 1 and 1.05 gives good results. Robustness to
variation in internal Hamiltonian parameters such as chemical shifts
can also be put into the fitness function; however, for typical liquid
state linewidths the pulses found are more than sufficiently robust.
This feature can also be used for incoherent averaging in the
subsystem approach. The convergence of the search algorithm is
dramatically improved by the use of conjugate gradient techniques.

To minimize decoherence, it is desirable to minimize the pulse
duration. However, because we do not apriori know even approximately
what this duration should be, it would take many optimizations for
different durations to find the optimum.  Furthermore, the time optimal point may require unrealistically high powers or fast pulse variations.  Instead, we allow the
length of time for each step of the pulse to vary and use timestep
derivatives to converge quickly to a good pulse duration. A penalty
function is added to limit the total time of the pulse and push the optimization towards shorter pulse times.

For high fidelity experimental control, smooth and slowly varying
``spectrometer friendly" pulses are ideal, although the desired
criteria can be enforced in the search with more sophisticated
techniques \cite{Werschnik:2005a}. These techniques require defining
the pulse by many points. We have found a simple method to
systematically find sufficiently smooth pulses for large systems,
where defining pulses by many points is not feasible:

\begin{enumerate}
\item Find a high-fidelity
pulse using relatively long time steps, e.g. $20\ \mu$s.
\item Digitally smooth this pulse with shorter timesteps of, e.g.
$1\ \mu$s, making sure that the pulse bandwidth is within the system
limitations.
\item Use this smoothed version of the pulse as an initial guess for
the numerical optimization. There will be a small loss of fidelity
from the smoothing. Nevertheless, it will be a good starting point.
\item If necessary, repeat the smoothing and optimization procedure.
\end{enumerate}  
Empirically we have found that the re-optimization of the pulse in the
third step changes the pulse very little, and once the new optimized
pulse is determined, it is still sufficiently smooth. It is also necessary to ensure that the beginning and end of the pulse go to zero to ensure a smooth experimental turn-on/off. This can be
achieved with a penalty function that penalizes high powers at the
beginning and the end.

In combining the two methods of control described in this paper we
have discovered an additional method that can be used to reduce the
time neded to find pulses on larger systems. This method 
provides a practical way to find certain types of pulses only on large
systems, but it does not address scalability issues. If the system can
be decomposed into possibly overlapping subsystems such that for each
subsystem, the desired unitary operator factors into one acting on the
subsystem and another acting on the complement, then we can find good
pulses by defining the fitness function as a weighted sum of fitness
functions for each subsystem. The subsystems must be defined so that
the dominant dynamics of the system is captured. In particular, every
strong coupling must be internal to at least one of the subsystems. We
simulate each subsystem separately to reduce the simulation
complexity.   Unlike the scalable sub-system approach discussed in section IV, 
the optimal control subystems will cover the entire system and the pulse is expected to implement the desired unitary 
on the full system without pre- or post-error terms. 

Take as an example the crotonic acid system
with seven qubits. The molecule is shown in Figure \ref{croto_table}.
Pulses implementing one-qubit rotations have the desired factorization
property with respect to any subsystem. To capture the strong
couplings, we use the following four overlapping subsystems:
$\{M,C_1\}$, $\{C_1,C_2\}$, $\{H_1, C_2, C_3, H_2\}$, and $\{C_3,
C_4\}$.  $C_1$ is covered by two subsystems capturing its strongest couplings.  If the pulse refocusses these strong couplings to $M$ and $C_2$ it is likely to also refocus weaker couplings to say $C_4$ without additional optimization.  In general though, if the pulse performs the desired subystem unitaries
with high fidelity there is no guarantee that the smaller unconsidered
couplings and/or many-body effects will not give a much lower fidelity
implementation on the full system. Empirically however, we have found
the constructed pulses work well on the full system, both for short
one-qubit gates and for longer coupling gates. Furthermore, it
substantially speeds up finding pulses for the seven qubit system. The
pulse shown if Figure \ref{GRAPEpulse} was found with this method. It
has 99.9 \% fidelity averaged over the above subsystems and showed a
99.7 \% fidelity on the entire seven qubit system.
  
\section{Pulse compiler}
\label{pulsecompiler_appendix}
The efficient pre- and post-error term analysis described in section
\ref{compilercontrol} contains two steps: the first is to
\emph{efficiently} compute the relevant dynamics of the system under
the RF pulse; the second is to extract optimal pre- and post-error
terms that are correctable and give a good representation of the
simulated pulse.\\

\emph{Capture the relevant dynamics:} Consider an illustrative three
spin system. Spins 1 and 2 are of the same species, whereas spin 3 is
different. The propagator $U$ of a pulse affecting spins 1 and 2 can
be written as
\begin{multline}
U = \exp\{-i\Delta t (\omega_{nut}\left(\cos\theta\left(X_1 +
 X_2\right) + \sin\theta\left(Y_1 + Y_2\right)\right) \\ + \omega_1
 Z_1 + \omega_2 Z_2 + \omega_3Z_3 + J_{12}Z_1Z_2 + J_{23} Z_2Z_3 +
 J_{13} Z_1Z_3)\},
 \end{multline}
where the control and internal Hamiltonian terms are defined as above.
If the control fields are functions of time, the pulse must be
discretized and integrated over many time steps.

The dynamics of the system can be expanded into contributions of
different coupling order. If coupling effects are ignored, then the
zeroth-order $U_0$ can be computed by three independent single spin
simulations as
\begin{multline}
U_0 = e^{-i\Delta t\left( \omega_{nut}\left(\cos\theta X_1 + \sin\theta Y_1\right) +  c_1 Z_1\right)} \\
 e^{-i\Delta t\left(  \omega_{nut}\left(\cos\theta X_2 + \sin\theta Y_2\right)+  c_2 Z_2\right)}e^{-i\Delta t\left( c_3 Z_3\right)}.
\end{multline}

The effects of the coupling can then be added in a perturbative
manner. The first step is to consider the effect of the couplings
between pairs of spins or subsystems: couplings 1-2, 2-3 and 1-3 are
considered. By considering only pairs of spins, coherent indirect
coupling effects are lost. For example, there is an extra effect on
the dynamics of qubit 1 from the coupling to qubit 2 because qubit 2
is in turn affected by the coupling to qubit 3. These higher order
coupling effects can be neglected when coupling effects during the
pulse are small. Moreover, for soft selective pulses in the weak
coupling approximation, many of the second-order contributions vanish
because all the internal Hamiltonians of the off-resonant spins commute
with each other, so that in most cases, stopping at first order is
sufficient to encompass all the relevant dynamics.

If the pulse is found using optimal control theory rather than a
pre-defined simple pulse shape, it may drive the system through a
nonintuitive path that makes use of indirect coupling dynamics to
achieve the desired unitary. Therefore we need to
perform a set of simulations between the subsystem defined for the
optimal control and each of the other spins. For example, a GRAPE
pulse may be found by considering only the carbon subsystem.
Each pairwise simulation encompasses the carbon spins and one of the
proton spins. The computational resources required are
larger than for a simple spin pairwise simulation, but they still scale
polynomially with the total register size, provided the subsystems
are of constant size. \\

\emph{Search for the optimal error terms:} From the efficient
simulations of the previous paragraph, we need to determine the pre-
and post-error terms surrounding the ideal operation. This is achieved
through a simple numerical search optimizing the following fitness
function:
\begin{equation}
\Phi = Real\left\{tr\left(U_{sim}^\dagger\left(U_{post}U_{ideal}U_{pre}\right)\right)\right\}/N.
\end{equation}
This compares the simulated unitary to its representation as the ideal
unitary, restricted to the relevant qubits, sandwiched between the
pre- and post-error terms. This fitness function is sensitive to
global phase because we are considering only a subsystem, and so to be
consistent when representing the gates on the full system, local
phases matter. To extract the optimal error terms one can use a
classical search algorithm. Note that this search is performed on a
relatively low-dimensional parameter space. For pulses designed for
one qubit, the single spin simulations require only two parameters for
the phases in the error terms on this spin. In the pairwise
simulations there are six parameters for the phases in the $Z$ and
$ZZ$ error terms on the coupled spins. 

\textit{Error term computation:} For standard shaped pulses designed
to act independently on one or more spins, the following procedure
allows one to compute error terms for the pulse
representation. 

\begin{enumerate}

\item Perform single spin simulations including the RF control
fields for each spin and optimize the Z error terms (with respect to
the intended evolution on the spin) to capture the effects of the
chemical shift and the transient Bloch-Siegert effect.

\item Perform simulations of each pair of spins, including the
pairwise coupling and the RF control fields. Optimize the Z and ZZ
error terms with respect to the intended evolution on both spins.
Remove the contribution of the Z error terms found in the first step
from the ones found here (by dividing the terms or subtracting the
exponents) to account for the fact that the simulation also accounts
for the effects modeled in the first step. Call these the
``incremental Z error terms''.

\item Determine the Z error terms for each spin as the product of the terms
found in the first step and the incremental terms from each
pair-wise simulation involving this spin. The ZZ error terms
are the ones found in the second step.

\end{enumerate}

If the pulse is designed to have an independent action on a set of
subsystems, the procedure needs to be modified by
replacing the single-spin and pairwise simulations with
single-subsystem and pairs-of-subsystems simulations. This requires
optimizing both Z and ZZ error terms for the single-subsystem
simulations, and determining incremental contributions for the pair
simulations by removing the contributions already obtained for spins
and pairs internal to each subsystem.

\section{Sequence optimizer}
\label{sequencecompiler_appendix}

The sequence compiler was built to automate, as much as possible, the
design of pulse sequences for an NMR QIP. The compiler takes as one input
the fixed information about the internal Hamiltonian, namely the
chemical shifts and J coupling values, and the error term information
for the pulses obtained by optimal control and implementing the set of
one-qubit rotations needed. The second input is a representation of
a goal sequence of quantum evolutions that implements the desired
algorithm or quantum network. The sequence compiler then determines
delays between pulses so as to minimize the total error of the
implementation compared to the intended evolution. The goal sequence
is described with a purpose built language. Here is an example of such
a sequence:

\begin{verbatim}
;pulse C190 0 @C1:X+
;zz 0.25 C1 C2
;refocus C3180 0.25
;pulse C290 0.75 @C2:0+
;z 0.5 C3
\end{verbatim}

This sequence requests a 90 degree rotation about the x-axis on $C_1$
followed by a ZZ90 (equivalent to a controlled Z gate up to single
qubit Z rotations) coupling gate between $C_1$ and $C_2$ followed by a
90 degree one-qubit gate on $C_2$ and a final Z rotation on $C_3$.
During the coupling period a ``floating'' refocussing pulse of 180
degrees about the y-axis is executed on C3. The @ instructions specify
state assumptions that may simplify the optimization. The output of
the compiler is a pulse sequence that can be directly executed by the
spectrometer. The compiler automates a number of the tasks discussed
in section \ref{compilercontrol}.

\begin{itemize}
\item{Phase tracking:} To avoid having to use one physical
spectrometer channel per qubit, the compiler tracks the evolution of
the nuclear rotating frames and the spectrometer channel frames
throughout the computation. When a pulse about a certain axis is
required for a nucleus, a simple calculation of the phase difference
between the channel frame and the nuclear frame determines the phase
at which the pulse should be sent in the channel frame to achieve a
specific phase in the target nuclear frame. As noted above this
freedom allows us to avoid having to refocus the chemical shifts at
every step. It also allows free (pulse-less) implementation of Z rotations by
executing a frame change on the target nuclei and updating the phases
of all subsequent pulses. In addition, the relative phase evolution of
the nuclei and the observation channel are tracked and the observation
phase is adjusted appropriately. The frequency changes can be
implemented by either changing the transmitter frequency or phase
ramping the pulse. The phase tracking calculations assume that the
frequency change is phase coherent, but experimentally this is not
always the case (particularly for large frequency changes), and we have
achieved more consistent results with phase ramping on our
spectrometer.

\item{State assumptions:} In some algorithms the state of the system
might be known at certain steps. For example, at the beginning of the
algorithm we may know that a particular qubit is in the state
$\ket{0}$ or the maximally mixed state $\openone$. This can simplify
the refocussing and phase tracking calculations. For example, if two
spins are in the maximally mixed state, the coupling between them has
no effect and so does not need to be refocussed. If a qubit is in a
pseudo-pure state then its coupling effect with the other spins is
reduced to an additional Z rotation, which can be accounted for with
the phase tracking calculation. State assumptions are implemented by
specifying the nuclear states after pulse commands, as shown in the
example. The compiler also implements some elementary state update
rules.

\item{Cancellation of \textit{virtual180s}:} A refocussing pulse
refocusses the couplings as desired but also affects the computation.
As discussed in the text, conventionally this has been avoided by
placing refocussing pulses in pairs to ensure that the state of the
refocussed spin is not changed. However, this second refocussing pulse
can be considered a virtual 180 that does not have to be implemented
physically. As such it can be implemented by absorbing it into the
next operation, modifying this operation appropriately. Using the notation $R_\phi(\theta)$ to denote a
rotation of $\theta$ about the axis an angle $\phi$ away from the
x-axis in the x-y plane, the following rotations are equivalent:
\begin{equation}
R_\alpha \left(\frac{\pi}{2}\right) R_\beta\left(\pi\right) = R_z\left(\gamma\right) R_\delta \left(\frac{\pi}{2}\right),
\end{equation}
where we use matrix multiplication and trigonometric identities to determine $\gamma$ and $\delta$:
\begin{align}
\gamma &= 2(\alpha-\beta) \\
\delta &= 2\beta-\alpha-\pi.
\end{align}
That is, a 180 degree pulse at phase $\beta$ followed by a 90
degree pulse at phase $\alpha$ is equivalent to a 90 degree pulse at
phase $2\beta-\alpha-\pi$ followed by a rotation about the z axis of
$2(\alpha-\beta)$. Since the z rotation comes for free as a frame
change, we have eliminated the need to do the second 180. The compiler
keeps track of these \textit{virtual180s}, which considerably
simplifies writing pulse programs. The \textit{virtual180} can also
be sent backwards through the pulse program in a similar manner and
absorbed into a preceding pulse. In some cases this may help the
refocussing scheme.

\item{Optimization of delays:} The pulse sequence can be considered as
a sequence of events (computation and refocussing pulses) with delays
in between. The delays between pulses serve the dual purposes of
allowing desired coupling gates to occur and also allowing time for
refocussing unwanted coupling effects. The refocussing pulses change
the direction in which the couplings are evolving, which can help reach the
coupling goals. At the beginning of each event certain couplings must
be at their goal values. In most cases, only those couplings that do
not commute with the pulse's intended effect have a fixed goal target.
Other couplings are simply tracked until a fixed goal is required
\cite{bowdrey:032315}. The coupling evolution for each pair of spins
is calculated at each event from both the coupling evolution during
the delays and the coupling errors terms in the pulse representations.
For those pairs that have a fixed target at this event the calculated
coupling is compared to the goal. A euclidian distance function is defined as the sum squared error between the goal and actual couplings and is related to an estimate of fidelity loss for the whole
sequence. Optimizing the pulse sequence is now reduced to the task of
optimizing the delays between each period. This is handled by a
simple iterative optimization to minimize the total contribution to
the distance function of the events bounding the delays. That is, the
delays are individually optimized one by one, starting at the last
one. After the first one is optimized, the sequential optimization
starts again at the last one, repeating the process until the
improvement is smaller than a threshold, or a goal distance is
achieved. 
Although effective, this optimization strategy is simplistic and easily gets trapped in local minima.  It would be useful to develop strategies that optimize all delays together with a non-linear least-squares optimization and consider different distance functions such as maximal error.  
\end{itemize}

\end{document}